# First Principles Reactive Flux Theory for Surface Reactions: Multiple Channels and Recrossing Dynamics

Chen Li,[1] Xiongzhi Zeng,[1] Yongle Li,[2] Zhenyu Li,[1] Hua Guo,[3] and Bin Jiang[1,*]

[1]State Key Laboratory of Precision and Intelligent Chemistry, Department of Chemical Physics, University of Science and Technology of China, Hefei, Anhui 230026, China

[2]Department of Physics, International Center of Quantum and Molecular Structures and Shanghai Key Laboratory of High Temperature Superconductors, Shanghai University, Shanghai 200444, China

[3]Department of Chemistry and Chemical Biology, Center for Computational Chemistry, University of New Mexico, Albuquerque, New Mexico 87131, USA

*: corresponding author: *bjiangch@ustc.edu.cn*

**Abstract**

Heterogenous reactions typically consist of multiple elementary steps and their rate coefficients are of fundamental importance in elucidating the mechanisms and micro-kinetics of these processes. Transition-state theory (TST) for calculating surface reaction rate coefficients often relies solely on the harmonic approximation of adsorbent vibrations and neglects recrossing dynamics. Here, we combine, for the first time, an efficient metadynamics enhanced sampling method with a more general reactive flux approach to calculate rate coefficients of surface reactions of any order and/or with multiple reaction coordinates, overcoming these limitations of TST. We apply this approach to a textbook surface reaction, CO oxidation on Pt(111), for which rate constants have been precisely measured, using a full-dimensional neural network potential energy surface constructed from first-principles data. An accurate multi-dimensional free-energy surface is obtained by incorporating three collective variables, yielding rate coefficients for both CO oxidation and the competing CO desorption that are in good agreement with experimental data. Interestingly, our results reveal significant dynamic recrossing in both channels, which however arises from distinct physical mechanisms. This approach represents an accurate and general framework for calculating rate coefficients of elementary surface processes from first-principles, which is vital for developing predictive kinetic models for heterogenous catalysis.

**Introduction**

Because of the pivotal role played by heterogeneous catalysis in modern society, it is highly desirable to improve efficiency and limit environmental impact of such processes. Fundamental knowledge of the mechanism and kinetics is the key to achieving these goals. A catalytic process typically consists of multiple elemental steps and the overall rate is determined by a network of micro-kinetic equations. A major challenge in understanding catalysis is the lack of theoretical methods for accurately predicting the rates of elementary surface processes.[1-4] Unlike gas phase reactions, interfacial processes are often affected by additional energy dissipation channels and inter-adsorbate interactions, which make such endeavors significantly more challenging.

Recent years have witnessed rapid progress in the experimental measurement of reaction rates for surface processes. For example, Wodtke and coworkers developed an accurate method based on velocity map imaging[5] and applied this so-called velocity resolved kinetics (VRK) method to accurately measure rate coefficients of several representative surface reactions.[6-12] These experimental rate measurements provide valuable benchmarks for testing rate theories.

Theoretically, the rate coefficient is typically estimated using the conventional transition-state theory (TST).[13, 14] TST is based on the premise that the thermal rate for a reaction is determined by the Boltzmann population at the activated complex, which can thus be calculated using statistical mechanics. As such, recrossing dynamics at the dividing surface separating reactants and products are neglected. To reduce errors introduced by the neglect of recrossing, the dividing surface can be chosen variationally to minimize the TST rate coefficient.[15] Furthermore, tunneling is not included in the TST framework and corrections are thus needed via various approximated methods.[15] For most TST calculations, the harmonic oscillator approximation is imposed in computing partition functions and one requires only the energetics and harmonic frequencies of the saddle point and reactant state along the minimum energy path on the potential energy surface (PES). For surface processes, such information can be readily obtained from density functional theory (DFT) calculations.[1, 2] While very efficient, harmonic TST neglects anharmonicity, which can have a significant impact on entropy. Such errors are particularly pronounced for soft modes such as adsorbate diffusion, where surface corrugation renders the PES particularly anharmonic.[16-18] Obviously, these approximations of TST may lead to significant errors in the calculated rate coefficient. Indeed, for CO oxidation on Pt(111), an industrially and fundamentally important surface reaction, the harmonic TST rate coefficients based on DFT data are 20–100 times larger than measured ones.[7] This large discrepancy underscores the need for a more accurate rate theory.

Beyond these TST shortcomings, the calculated rate coefficient may also suffer from inaccuracies of the generalized gradient approximation (GGA) density functionals commonly used in DFT calculations,[19] which can affect both the enthalpic and entropic contributions. To delineate the origin of errors in rate calculations, it is thus important to compare experimental data with accurate rate theory, which can provide a quantitative assessment of the accuracy of the functional used in generating the PES.

The harmonic approximation in TST can be avoided by computing the free-energy barrier of an elementary process by molecular dynamics (MD), followed by estimating the rate coefficient via statistical mechanics. Indeed, such calculations have been reported for several surface processes using either ab initio MD (AIMD) or MD on machine learned

PESs,[20-29] with various enhanced sampling techniques.[30-32] However, these studies typically involve a single-channel process where one reaction coordinate (RC) is often sufficient. With few exceptions,[33] the influence of dynamical recrossing is ignored.

A more general and rigorous approach for computing rate coefficients of reactions is the reactive flux theory (RFT), which expresses the rate coefficient in terms of a flux-side correlation function.[34, 35] Further factorization via the Bennett-Chandler method[36, 37] results in two terms, a TST rate coefficient in the short-time limit and a transmission coefficient accounting for the recrossing effects, both of which can be accurately determined by MD.[38] In the gas phase, RFT has been combined with the path-integral based ring polymer molecular dynamics (RPMD) method[39] to include nuclear quantum effects such as tunneling and zero-point energy.[40-44] RPMD replaces a quantum particle with a necklace of harmonically connected beads and as a result the dynamics can be followed by classical motion of the ring polymer. This classical-quantum isomorphism enables the use of Newton's equation in simulating the dynamics, thus numerically efficient. Using ab initio PESs, the RPMD rate theory has been validated by comparing with exact quantum dynamics calculations for several gaseous bimolecular reactions.[45] Applications of the RPMD rate theory to surface diffusion have also been reported.[46, 47] Very recently, we have presented RPMD calculations of rates for NO desorption from Pd(111)[48] and for $H_2$ dissociation/recombination on multiple metal surfaces.[49-51] Agreement with experimental results is quite satisfactory. Nevertheless, these RFT calculations so far have been restricted to processes involving a single specific RC, in which the umbrella sampling method has been used to evaluate the one-dimensional free-energy profile.[52] On surfaces, several reaction channels may coexist with similar barriers, and these competing processes might be temporally and spatially inseparable. In such cases, more than one RCs are needed in rate coefficient calculations and a more efficient enhanced sampling method is necessary.

In this work, we discuss a general expression for rate coefficients of surface reactions of any order within the RFT. The RFT is implemented using a well-tempered metadynamics (WT-MetaD)[53] method to treat multiple RCs and applied to CO oxidation on Pt(111), which is not only an important reaction involved in many industrially catalytic processes[54] but also a prototype for testing new theories/experiments.[7, 55, 56] To make it tractable, we take advantage of a high-dimensional neural network PES,[57, 58] which possesses a comparable accuracy to DFT energy but with orders of magnitude higher efficiency. This PES allows the inclusion of both the molecular and surface degrees of freedom (DOFs) and efficient exploration of the multi-dimensional free-energy surface (FES) by WT-MetaD. With no adjustable parameters, the calculated rate coefficients of both the CO oxidation and CO desorption processes are in reasonably good agreement with the experiment. Interestingly, significant recrossing effects are uncovered in both channels, originated from distinct physical mechanisms, highlighting the importance of dynamics in surface reaction kinetics. These theoretical results shed valuable light on the microscopic details of surface processes.

**Results**

As detailed in the Methods section, the RFT used in this work overcomes the limitations of conventional TST for surface reactions in several aspects. First, the multi-dimensional free-energy surface is explored by an enhanced sampling method—WT-MetaD—which fully accounts for the anharmonicity of vibrational contributions to the potential of mean force (PMF) along the RC. Second, surface atoms are allowed to move and their soft vibrational modes are included in free-energy calculations. Third, the recrossing factor is determined by lunching numerous trajectories at the dividing surface towards both the reactant and product sides and counting those returned ones with opposite velocities. Apparently, such extensive MD simulations are far beyond the limit of AIMD and constructing an analytical PES becomes essential. To this end, the embedded atom neural network (EANN) method is used to fit thousands of DFT data points to yield the high-dimensional PES with first-principles accuracy, as described in Methods and Supplementary Information (SI).

**Potential energy surface**

Fig. 1 shows the energy profile from CO co-adsorption with O, CO oxidation, to $CO_2$ desorption on Pt(111). The geometries of the relevant stationary points are shown in Table S1 in SI. Consistent with previous theoretical studies,[18, 59-61] an isolated CO molecule is found to adsorb perpendicularly on an fcc site with C-terminal facing down. The adsorption energy of 1.42 eV is in good agreement with the experimental value of 1.4 eV.[62] When the CO molecule and O atom co-adsorb on the Pt(111) surface, the desorption energy of CO becomes 1.30 eV, which may be caused by the repulsion of the adsorbed O atom. The O atom also prefers an fcc site. The adsorbed CO can easily diffuse from the fcc site to the bridge site by overcoming a small potential barrier of ~0.05 eV. In contrast, the diffusion of the O atom on the Pt(111) surface is more difficult for its relatively high diffusion barrier (0.57 eV).

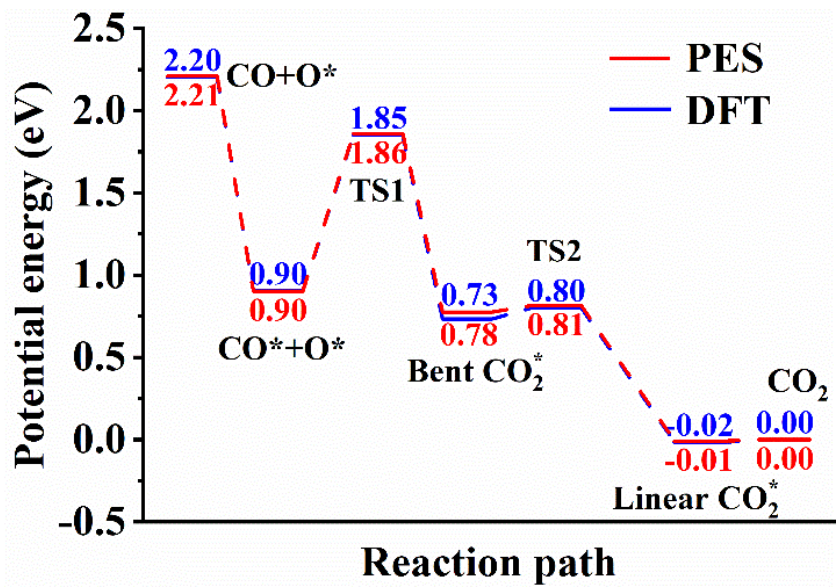

**Fig. 1. Potential energy profile for CO oxidation/desorption on Pt(111).** Comparison of energetics the relevant stationary points predicted by the EANN PES and DFT.

At the transition state (TS1) of CO oxidation, the adsorbed CO* molecule approaches the adsorbed O atom (O*), forming a C*-O* bond above a hcp site with the bond length of 1.96 Å and an O-C*-O* angle of 110°, while the spectator C*-O moiety has a bond length of 1.17 Å with the O end pointing outward at the top site. This step proceeds with the barrier of 0.95 eV. Both the geometry and barrier calculated here are in good agreement with the predictions of the PBE[20] and PW91[56] functionals. However, the barrier is slightly lower than 1.06 eV reported with a 2×2 supercell,[18] and slightly higher than 0.8 eV obtained with a van der Waals corrected RPBE functional.[7] After sliding from the TS1 to the product side, the subsequent elementary step is the transformation of the bent chemisorbed $CO_2$ to the linear physisorbed $CO_2$, in which the transition state (TS2) has both CO bonds with a length of about 1.2 Å with a bond angle of 147°. The potential barrier for this step is 0.07 eV, lower than that (0.21 eV) calculated by Zhou et al. using the PW91 functional.[56]

**Free energy surface**

For free-energy calculations, the choice of suitable collective variables (CVs) is of paramount importance for all CV-based enhanced sampling methods, including WT-MetaD.[53] During CO oxidation on Pt(111), adsorbed CO can be competitively consumed through desorption with a comparable barrier to the reaction. Consequently, these two elementary steps are not separable in space and time.[7] As discussed in detail in SI, the CO desorption channel cannot be excluded when sampling the phase space for CO oxidation in the WT-MetaD simulations. In addition, the indistinguishability of the two oxygen atoms results in two identical CO desorption channels. To describe this multi-channel surface process, as illustrated in Fig. S1, it is essential to introduce three CVs, including the permutationally equivalent distances between the C atom and each O atom ($r_1$ and $r_2$) and the height of the C atom above the top layer of the surface ($h$). As shown in Fig. S2, the calculated PMFs for CO oxidation deviate from the converged result when insufficient CVs are used, which fails to adequately sample the relevant phase space, leading to inaccuracies in the free-energy calculations. Notably, by using $h$ as one of the CVs, the CO desorption pathway is naturally incorporated in the enhanced sampling, allowing us to obtain the PMF and rate coefficients for this channel without additional effort.

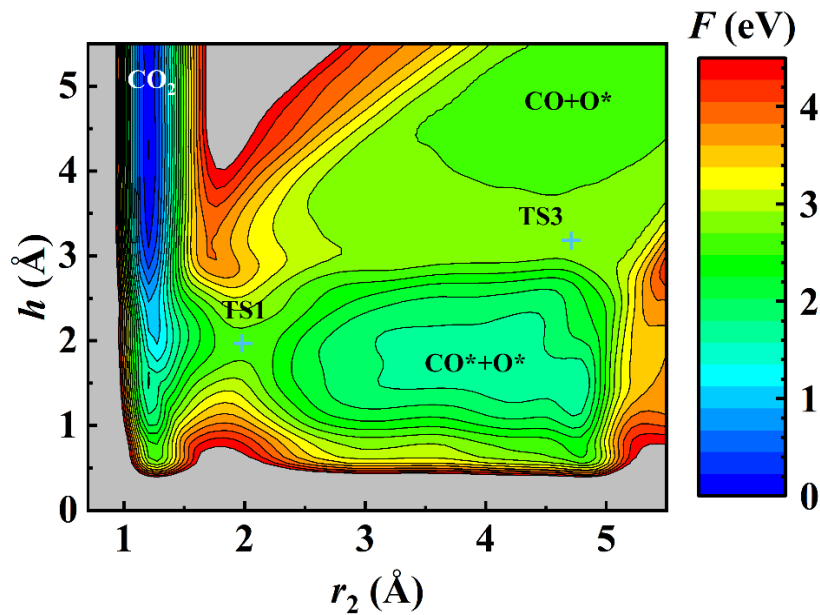

**Fig. 2. Two-dimensional FES for describing CO oxidation and CO desorption on Pt(111).** This FES is obtained by integrating over $r_1$ the three-dimensional FES, $F(r_1, r_2, h)$, which was computed by WT-MetaD method at 600 K. Note that there exists another permutationally symmetric CO desorption channel, which is not shown due to the integration over $r_1$.

Based on the newly constructed EANN PES, the three-dimensional FES, $F(r_1, r_2, h)$, was computed using WT-MetaD from 500 to 700 K at a 50 K interval. For visualization, we integrate over $r_1$ to obtain the two-dimensional FES at 600 K, shown in Fig. 2. As can be seen, the initial state (OC* + O*) accesses the reaction channel through TS1 largely along the $r_2$ coordinate, leading to the production of gaseous $CO_2$. The free-energy barrier (TS1) is 1.02 eV. On the other hand, the CO desorption channel also has a free-energy barrier TS3 (1.11 eV). Interestingly, the existence of a small exit channel barrier for CO desorption is consistent with the earlier work of Hu and coworkers.[22]

**Recrossing dynamics**

Fig. 3 shows the time-dependent transmission coefficients for the CO oxidation reaction and CO desorption at 500, 600, and 700 K. For the reaction channel (Fig. 3a), the transmission coefficients are calculated by placing the dividing surface at the TS1 ($r_2$ = 1.98 Å). Interestingly, somewhat similar to the desorption of NO on the frozen Pd(111) surface,[48] $\kappa$ decreases rapidly after experiencing a short plateau at the beginning. Its value finally converges to ~ 0.35 at 50 ps, indicating that the recrossing effect is significant at TS1. As shown by an exemplary trajectory in SI (Fig. S3a), the recrossing can be largely attributed to the inefficient energy transfer from the C-O vibrational DOF, which is aligned with the RC at TS1, to the $CO_2$ translational DOF along the $Z$ coordinate, in order for it to escape.

In Fig. 3b, the transmission coefficients for the desorption channel, are computed by placing the dividing surface at the TS3 ($h$ = 3.04 Å). The recrossing effect of the CO desorption on Pt(111) is much more pronounced than desorption of NO on Pd(111),[48] with $\kappa$ converging to about 0.3 in less than 1 ps. The significant recrossing effect here is probably due to the rebound of CO from the surface to recross the barrier, as illustrated by an exemplary recrossing trajectory in SI (Fig. S4a).

Interestingly, the CO oxidation and desorption channels are dominated by two different types of recrossing behaviors. In the CO oxidation channel, major recrossing trajectories start with positive velocities towards the product side ($CO_2$) and recross the TS1 to arrive the reactant side (CO*+O*). While in the CO desorption case, most recrossing trajectories correspond to those with initial negative velocities towards the CO adsorption state and then recross the TS3 after the rebound from the surface to yield the desorbed CO products. Both behaviors lead to small transmission coefficients.

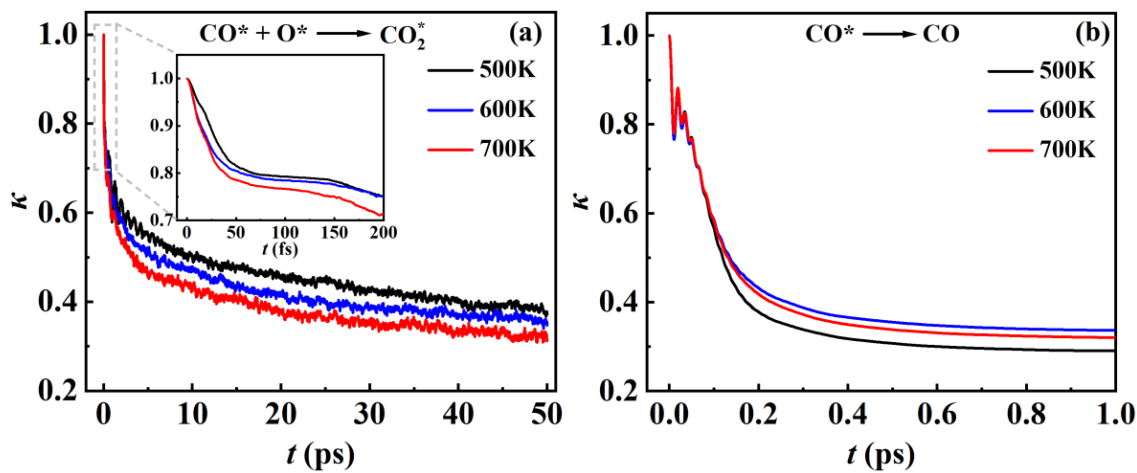

**Fig. 3. Time evolution of the transmission coefficients. a** and **b** show the transmission coefficients for the oxidation and desorption of CO on Pt(111) at 500, 600 and 700 K, respectively.

**Comparison with experimental data**

In Fig. 4a, the rate coefficients for CO oxidation on Pt(111) predicted by RFT are compared with the recent experimental data.[7] The agreement is quite satisfactory, especially given the fact that no adjustable parameters are used. In addition, two sets of TST results (HA and HT) are also included in the figure. The RFT predictions are obviously in better agreement with the experiment than both TST calculations. As expected, the HA treatment of the translational DOFs of the adsorbates on the surface overestimates the rate coefficients presumably because it underestimates the OC* diffusional entropy, consistent with the conclusion of the previous discussion.[5] This problem can be improved by the HT model,[16, 17] which considers the two-dimensional surface corrugation in the translational entropy estimate and produces a more reasonable pre-factor. Note that the larger discrepancy between the HA model and experiment reported in Ref. 7 likely stems from both the inaccurate energy barrier (~0.21 eV lower than our result) and the harmonic approximation of vibrational frequencies[63]. The much better agreement between the RFT results and experiment underscores the importance of free-energy simulation of the reaction path and the recrossing dynamics. We note that the RFT rate coefficients are still 2-3 times larger than the experiment ones. There might be several possible reasons for the discrepancy, such as inaccuracies in the GGA functional used in the DFT calculation of the PES and the relatively small supercell size.

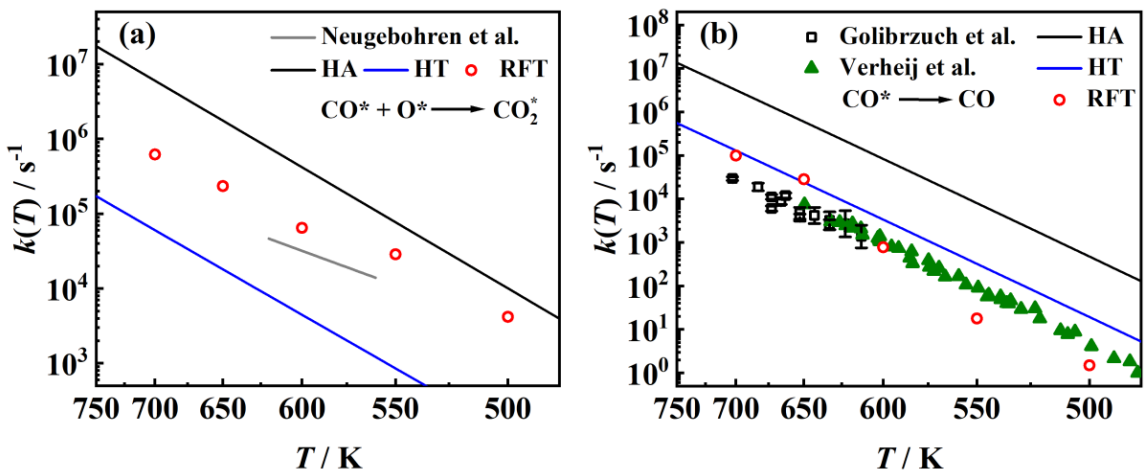

**Fig. 4. CO oxidation and desorption rate coefficients.** Comparison of the RFT rate coefficient and the experimental data for CO oxidation **a** and desorption **b**. HA and HT show the rate coefficients by TST with the harmonic approximation and hindered translator model, respectively. The experimental CO oxidation rate coefficients are those of

Neugebohren et al.[7] The CO desorption rate coefficients are from Golibrzuch et al.[6] and Verheij et al.[64]

In Fig. 4b, the CO desorption rate coefficients from Pt(111) calculated from RFT are compared with the available experimental data.[6, 64] As discussed above, this is an added bonus as the FES naturally describes the desorption channel. The theory-experiment agreement is semi-quantitative. Not surprisingly, the HA model again overestimates the rate coefficients by several orders of magnitude due presumably to the underestimation of the adsorbed CO diffusion entropy. Different from the CO oxidation reaction discussed above, however, the absolute values of the HT and RFT rate coefficients for CO desorption are both close to the experimental results. We note that the experimental CO desorption rates were measured in the absence of co-adsorbed oxygen,[6,64] while our calculations adapted a model with co-adsorbed oxygen. Our DFT calculations have shown that the co-adsorption leads to a 0.1 eV difference of adsorption energy, which is expected to influence the theory-experiment agreement. Second and perhaps more importantly, it is well known that GGA density functionals have trouble in reproducing the experimentally determined preferred CO adsorption sites on platinum group metal surfaces, due to self-interaction errors.[65] This problem is expected to affect the topography of the PES and the partition functions of adsorbed CO. For these reasons, our current model in characterizing CO desorption is much less reliable than that for CO oxidation. On the other hand, since the HT-TST model relies only on the DFT-optimized stationary point properties of the free and adsorbed CO molecules and ignores the free-energy landscape, its reproduction of the experimental activation energy and pre-factor might also be accidental. Overall, we feel the level of the experiment-theory agreement achieved here for this secondary channel is quite satisfactory. However, further investigation about the effect of PES on the CO desorption from Pt(111) is certainly warranted.

**Discussion**

Surface reaction rate coefficients are of fundamental importance for heterogeneous processes. However, the accuracy of the commonly used transition state theory for calculating surface reaction rate coefficients is significantly limited by the harmonic approximation and ignorance of the recrossing dynamics. Fortunately, these shortcomings can be overcome in the reactive flux theory. But thus far, most RFT calculations on surface reactions have been restricted to processes involving a single reaction coordiante. Indeed, on surfaces, several reaction channels may coexist with similar barriers and may be inseparable, such as the system discussed in this work. Herein, we present a general reactive flux theory for calculating rate coefficients of surface reactions of any-order with multiple channels. In particular, the rate coefficient is factorized into static and dynamic contributions, both computed using classical molecular dynamics. The former is enabled by the multi-dimensional free-energy surface explored by an efficient well-tempered metadynamics method that can incorporate multiple reaction coordinates, while the latter is simulated by recrossing trajectories initiated at the dividing surface at the transition state.

This theory is applied to a prototypical surface reaction, namely CO oxidation on Pt(111), for which experimental data are available. The simulation is accelerated by a high-dimensional neural network potential energy surface for the (CO+O)/Pt(111) system fitted to first-principles data, which is orders of magnitude more efficient than direct dynamics. A complication in this system is that the CO oxidation reaction is in direct competition with CO desorption, which demands more than a single reaction coordinate. To this end, we utilize three collective variables to construct a multi-dimensional free-energy surface covering all the above elementary steps.

The calculated CO oxidation rate coefficients are found in very good agreement with experiment. This success not only validates the reactive flux theory, but also exposes the failure of conventional transition-state theory in quantitatively evaluating rate coefficients. Our calculated CO desorption rates are also in reasonably good agreement with experiment, although for this channel our model contains more uncertainties than for the CO oxidation. It suggests that the GGA DFT is reasonably accurate for the CO oxidation process, despite possible inaccuracies for the CO desorption channel. Our calculations further reveal significant recrossing dynamics and identify different origins of such dynamical effects. Overall, our results demonstrate the advantage of combining reactive flux rate theory with advanced simulation techniques to simultaneously calculate multiple elementary surface reaction rates in complex systems from first principles, laying the foundation for accurate micro-kinetic modeling of heterogeneous catalysis.

## Methods

### Transition-state theory

In the traditional TST,[13, 14] the rate coefficient for a reaction is simply expressed as,

$$k(T) = \frac{k_B T}{h} \frac{Q^{\ddagger'}}{Q_r} \exp\left(-\frac{E_0}{k_B T}\right), \quad (1)$$

where $E_0$ is the zero-point energy corrected potential barrier, $Q^{\ddagger'}$ and $Q_r$ are partition functions for the transition state (TS) excluding the RC and for the reactant(s). $k_B$ and $h$ are the Boltzmann and Planck constants, respectively. Eq. (1) is often used within the harmonic approximation, by which $E_0$ and partition functions can be determined by energies and harmonic vibrational frequencies of the TS and reactant(s) obtained by a first principles method, such as GGA-DFT.

On the other hand, according to the relationship between equilibrium constant and free-energy change, the TST rate coefficient can also be expressed as:[38]

$$k(T) = \frac{k_B T}{h} \exp\left(-\frac{\Delta F^\ddagger}{k_B T}\right) \qquad (2)$$

where the free-energy change from the reactant to the TS, $\Delta F^\ddagger$, can be computed by MD with enhanced sampling. The MD calculation includes anharmonicity, but there is no easy way to calculate the recrossing effect in the traditional TST framework.

**Reactive flux theory**

The RFT approach of Miller and coworkers[35] defines the reaction rate coefficient in terms of the flux-side correlation function at a dividing surface separating the reactant and the product, $c_{fs}(t)$:

$$k(T) = \frac{c_{fs}(t \to \infty)}{Q_r}, \qquad (3)$$

where $Q_r$ is the reactant partition function. The actual computation can be simplified by the Bennett-Chandler factorization,[36, 37]

$$k(T) = \kappa(t \to \infty) k_{TST}(T). \qquad (4)$$

Here, $\kappa(t\to\infty)=c_{fs}(t\to\infty)/c_{fs}(t\to 0^+)$ is the long-time limit of the time-dependent transmission coefficient, providing a quantitative way of evaluating the recrossing effect at the dividing surface. Specifically, it can be calculated by sampling a constrained ensemble at the dividing surface to estimate the ratio of the two flux-side correlation functions,[42] namely,

$$\kappa(t\to\infty)=\frac{\langle\delta[s(\mathbf{x})]v_s(\mathbf{p},\mathbf{x})h[s(\mathbf{x}_t)]\rangle}{\langle\delta[s(\mathbf{x})]v_s(\mathbf{p},\mathbf{x})h[v_s(\mathbf{p},\mathbf{x})]\rangle}, \tag{5}$$

where $s(\mathbf{x})$ is the dividing surface defined in the RC, $\xi(\mathbf{x})$,

$$s(\mathbf{x})=\xi(\mathbf{x})-\xi^{\ddagger}. \tag{6}$$

Here, $\xi^{\ddagger}$ is the reference RC at the position of the dividing surface, often located at the barrier, and $v_s(\mathbf{p},\mathbf{x})$ is its time derivative or velocity, $\mathbf{x}$ and $\mathbf{p}$ denote the initial coordinates and momenta of the atoms in the system, respectively, $\mathbf{x}_t$ denotes the coordinates at time $t$.

On the other hand, $k_{\mathrm{TST}}(T)=c_{fs}(t\to 0^+)/Q_r(T)$ is the TST rate coefficient in the short-time limit, which accounts for the static contribution to the rate coefficient. This term needs to be evaluated by integrating the free energy from the reactant side ($\xi^{(0)}$) along the RC up to the dividing surface ($\xi^{\ddagger}$), by which the reactant partition function is fully taken into account. A general expression is given by,[38]

$$k_{\text{TST}}(T) = A \frac{e^{-\beta F(\xi^{\ddagger})}}{\int_{\xi^{(0)}}^{\xi^{\ddagger}} e^{-\beta F(\xi)} d\xi}. \tag{7}$$

Here, the preexponential factor $A$ is determined by the reactant partition function, which gives rise to the unit of the rate coefficient. For example, it is $1/(2\pi\beta m_{\xi})^{1/2}$ for a first-order reaction, $4\pi\left[(\xi^{\ddagger})^3 - (\xi^{(0)})^3\right]/3(2\pi\beta m_{\xi})^{1/2}$ or $\pi\left[(\xi^{\ddagger})^2 - (\xi^{(0)})^2\right]/(2\pi\beta m_{\xi})^{1/2}$ for a second-order reaction in the gas phase or on a surface, respectively, with $m_{\xi}$ and $F(\xi)$ being the reduced mass and free energy along $\xi(\mathbf{x})$. Note that the application of Eq. (7) requires the RC to be either cartesian or distance coordinates.[42]

A distinct benefit of the Bennett-Chandler factorization is that the final result becomes independent of the choice of the dividing surface.[41] In other words, a non-optimally chosen dividing surface is counterbalanced by the transmission coefficient. This property is highly desired in practice because an optimal choice of the dividing surface is nearly impossible in a multidimensional configuration space. It should be pointed out that the RPMD rate theory[40-44] is based on the same theoretical foundation, but replacing the classical Hamiltonian with its ring-polymer counterpart. As a result, the single-bead RPMD rate coefficient is identical to rate coefficient obtained using the classical RFT. For systems that nuclear quantum effects are not important, such as CO oxidation discussed here, the latter is sufficient.

**Well-tempered metadynamics**

For gas phase elementary reactions, it is often sufficient to define a single RC.[44, 45] For complex reactions such as those occurring on surfaces, however, multiple elementary processes might be coupled.[25] To deal with such problems, one often needs to introduce multiple collective variables to map out the multi-dimensional free-energy surface (FES),[66,67] which can be then projected along a specific RC to obtain the free-energy profile, known as the potential of mean force (PMF) for the reaction. For this purpose, the umbrella sampling technique[52] becomes numerically expensive. A more efficient enhanced sampling method is metadynamics,[68] in which Gaussian shaped potentials are added in the CVs to flatten the FES. In this work, we take advantage of the well-tempered version of metadynamics (WT-MetaD),[53] in which the dynamics is biased with adaptive history-dependent Gaussian potentials, namely,

$$V^{(b)}(\mathbf{CV}(\mathbf{x}),t) = \sum_{k\tau<t} W(k\tau) \exp\left(-\sum_{i=1}^{d} \frac{\left(CV_i(\mathbf{x}) - CV_i(\mathbf{x}(k\tau))\right)^2}{2\sigma_i^2}\right). \tag{8}$$

Compared with the standard metadynamics,[68] WT-MetaD adopts the history-dependent Gaussian height $W(k\tau)$ and converges more easily. Here, $d$ denotes the number of CVs and $\sigma_i$ is the Gaussian width corresponding to the $i$th CV. The final FES can be later obtained by removing the Gaussian potentials.

**Application to CO oxidation on Pt(111)**

As proof-of-concept, this general RFT approach is applied to CO oxidation on Pt(111). The PES for the CO + O system on Pt(111) was constructed by the embedded atom neural network (EANN) method.[69] The Pt(111) surface was modeled by a four-layer slab within a 3×3 surface supercell, in which the top two layers are mobile. DFT points were calculated with the revised Perdew–Burke–Ernzerhof (RPBE) functional.[70] In each supercell, there are two oxygen and one carbon atoms, plus eighteen movable Pt atoms. The details of the DFT calculation and PES training can be found in Supporting Information (SI).

**Computational details**

All MD calculations are performed using the LAMMPS package[71] supplemented by PLUMED v2.4 plugin.[72] The canonical ensemble from 500 to 700 K is sampled for 40 ns using a Nosé-Hoover thermostat and a time step of 0.2 fs. Regarding WT-MetaD, the initial Gaussian height ($W_0$) is set to 0.02 eV, the widths ($\sigma_i$) of the Gaussians added along each CV are all taken to be 0.2 Å, and the time interval ($\tau$) for adding Gaussians is 80 fs. The bias factor $(T+\Delta T)/\Delta T$ is chosen based on the maximum potential barrier for the elementary reactions. Finally, $k_{\text{TST}}(T)$ is obtained according to Eq. (7), following a one-dimensional (1D) PMF along the chosen RC for $CO_2$ formation ($r_1$ or $r_2$) or CO desorption ($h$) channel, after the multi-dimensional FES is integrated over other CVs. $\kappa(t \to \infty)$ is computed by sampling an ensemble of trajectories (say 20000) at the barrier of the corresponding 1D PMF and allowing them to evolve towards either side. These quantities are obtained using a heavily modified version of the RPMDrate program.[44] The computed free-energy profile has been carefully checked for convergence with regard to the sampling time.

For comparison, we also computed the rate coefficients of CO oxidation and desorption with the traditional TST based on two translational entropy models of the adsorbates, namely the harmonic approximation (HA) and the hindered translator (HT) models.[16, 17] One thing to note is that the degeneracy of the transition state of the CO oxidation reaction is 3 due to the fact that there are three equivalent approaches of CO* towards O* which occupies the fcc site. Moreover, the CO doses are so small in experiment that the O* coverage is independent of time during the reaction,[5, 7] for which the CO oxidation reaction can be treated as a pseudo $1^{st}$-order reaction. To facilitate comparison, we only report the $1^{st}$-order rate coefficients, as done in the experiment.

**Data availability:**

The training dataset and trained PES will be made publicly available upon acceptance of the manuscript.

**Code availability:**

LAMMPS, PLUMED v2.4 plugin and EANN are all open source and freely available. The heavily modified RPMDrate program are available from the corresponding author on request.

**References:**

1. Bruix A, Margraf JT, Andersen M, Reuter K. First-principles-based multiscale modelling of heterogeneous catalysis. *Nat. Catal.* **2**, 659-670 (2019).
2. Motagamwala AH, Dumesic JA. Microkinetic modeling: A tool for rational catalyst design. *Chem. Rev.* **121**, 1049-1076 (2021).
3. Xie W, Xu J, Chen J, Wang H, Hu P. Achieving theory–experiment parity for activity and selectivity in heterogeneous catalysis using microkinetic modeling. *Acc. Chem. Res.* **55**, 1237-1248 (2022).
4. Chen Z, Liu Z, Xu X. XPK: Toward accurate and efficient microkinetic modeling in heterogeneous catalysis. *ACS Catal.* **13**, 15219-15229 (2023).
5. Park GB, *et al.* The kinetics of elementary thermal reactions in heterogeneous catalysis. *Nat. Rev. Chem.* **3**, 723-732 (2019).
6. Golibrzuch K, *et al.* CO desorption from a catalytic surface: Elucidation of the role of steps by velocity-selected residence time measurements. *J. Am. Chem. Soc.* **137**, 1465-1475 (2015).
7. Neugebohren J, *et al.* Velocity-resolved kinetics of site-specific carbon monoxide oxidation on platinum surfaces. *Nature* **558**, 280-283 (2018).
8. Borodin D, *et al.* Measuring transient reaction rates from nonstationary catalysts. *ACS Catal.* **10**, 14056-14066 (2020).
9. Borodin D, *et al.* Kinetics of NH3 Desorption and Diffusion on Pt: Implications for the Ostwald Process. *J. Am. Chem. Soc.* **143**, 18305-18316 (2021).
10. Borodin D, *et al.* The puzzle of rapid hydrogen oxidation on Pt(111). *Mole. Phys.* **119**, e1966533 (2021).
11. Fingerhut J, *et al.* The barrier for $CO_2$ functionalization to formate on hydrogenated Pt. *J. Phys. Chem. A* **125**, 7396-7405 (2021).
12. Borodin D, *et al.* Quantum effects in thermal reaction rates at metal surfaces. *Science* **377**, 394-398 (2022).
13. Eyring H. The activated complex in chemical reactions. *J. Chem. Phys.* **3**, 107-115 (1935).
14. Evans MG, Polanyi M. Some applications of the transition state method to the calculation of reaction velocities, especially in solution. *Trans. Faraday Soc.* **31**, 875-894 (1935).
15. Truhlar DG, Garrett BC, Klippenstein SJ. Current status of transition-state theory. *J. Phys. Chem.* **100**, 12771-12800 (1996).
16. Campbell CT, Sprowl LH, Árnadóttir L. Equilibrium constants and rate constants for adsorbates: Two-dimensional (2D) ideal Gas, 2D ideal lattice gas, and ideal hindered translator models. *J. Phys. Chem. C* **120**, 10283-10297 (2016).
17. Sprowl LH, Campbell CT, Árnadóttir L. Hindered translator and hindered rotor models for adsorbates: Partition functions and entropies. *J. Phys. Chem. C* **120**, 9719-9731 (2016).
18. Jørgensen M, Grönbeck H. Adsorbate entropies with complete potential energy sampling in microkinetic modeling. *J. Phys. Chem. C* **121**, 7199-7207 (2017).
19. Kroes G-J. Computational approaches to dissociative chemisorption on metals: Towards chemical accuracy. *Phys. Chem. Chem. Phys.* **23**, 8962-9048 (2021).

20. Xu J, Huang H, Hu P. An approach to calculate the free energy changes of surface reactions using free energy decomposition on ab initio brute-force molecular dynamics trajectories. *Phys. Chem. Chem. Phys.* **22**, 21340-21349 (2020).

21. Chen Z-N, Shen L, Yang M, Fu G, Hu H. Enhanced Ab Initio Molecular Dynamics Simulation of the Temperature-Dependent Thermodynamics for the Diffusion of Carbon Monoxide on Ru(0001) Surface. *J. Phys. Chem. C* **119**, 26422-26428 (2015).

22. Guo C, Wang Z, Wang D, Wang H-F, Hu P. First-principles determination of CO adsorption and desorption on Pt(111) in the free energy landscape. *J. Phys. Chem. C* **122**, 21478-21483 (2018).

23. Zeng X, Qiu Z, Li P, Li Z, Yang J. Steric hindrance effect in high-temperature reactions. *CCS Chem.* **2**, 460-467 (2020).

24. Yang M, Raucci U, Parrinello M. Reactant-induced dynamics of lithium imide surfaces during the ammonia decomposition process. *Nat. Catal.* **6**, 829-836 (2023).

25. Bonati L, *et al.* The role of dynamics in heterogeneous catalysis: Surface diffusivity and $N_2$ decomposition on Fe(111). *Proc. Natl. Acad. Sci. U. S. A.* **120**, e2313023120 (2023).

26. Stocker S, Jung H, Csányi G, Goldsmith CF, Reuter K, Margraf JT. Estimating free energy barriers for heterogeneous catalytic reactions with machine learning potentials and umbrella integration. *J. Chem. Theo. Comput.* **19**, 6796-6804 (2023).

27. Hu Z-Y, Luo L-H, Shang C, Liu Z-P. Free energy pathway exploration of catalytic formic acid decomposition on Pt-group metals in aqueous surroundings. *ACS Catal.* **14**, 7684-7695 (2024).

28. Sun J-J, Cheng J. Solid-to-liquid phase transitions of sub-nanometer clusters enhance chemical transformation. *Nat. Commun.* **10**, 5400 (2019).

29. Gong F-Q, Liu Y-P, Wang Y, E W, Tian Z-Q, Cheng J. Machine Learning Molecular Dynamics Shows Anomalous Entropic Effect on Catalysis through Surface Pre-melting of Nanoclusters. *Angew. Chem. Inter. Ed.* **63**, e202405379 (2024).

30. Abrams C, Bussi G. Enhanced sampling in molecular dynamics using metadynamics, replica-exchange, and temperature-acceleration. *Entropy* **16**, 163-199 (2014).

31. Yang YI, Shao Q, Zhang J, Yang L, Gao YQ. Enhanced sampling in molecular dynamics. *J. Chem. Phys.* **151**, (2019).

32. Ray D, Parrinello M. Kinetics from metadynamics: Principles, applications, and outlook. *J. Chem. Theo. Comput.* **19**, 5649-5670 (2023).

33. Galparsoro O, Kaufmann S, Auerbach DJ, Kandratsenka A, Wodtke AM. First principles rates for surface chemistry employing exact transition state theory: application to recombinative desorption of hydrogen from Cu(111). *Phys. Chem. Chem. Phys.* **22**, 17532-17539 (2020).

34. Yamamoto T. Quantum statistical mechanical theory of the rate of exchange chemical reactions in the gas phase. *J. Chem. Phys.* **33**, 281-289 (1960).

35. Miller WH, Schwartz SD, Tromp JW. Quantum mechanical rate constants for bimolecular reactions. *J. Chem. Phys.* **79**, 4889-4899 (1983).

36. Bennett CH. Molecular dynamics and transition state theory: the simulation of infrequent events. In: *Algorithms for Chemical Computations, ACS Symposium Series* (ed Christofferson RE). ACS (1977).

37. Chandler D. Statistical mechanics of isomerization dynamics in liquids and the transition state approximation. *J. Chem. Phys.* **68**, 2959-2970 (1978).

38. Peters B. Reactive Flux. In: *Reaction Rate Theory and Rare Events Simulations* (ed Peters B). Elsevier (2017).
39. Habershon S, Manolopoulos DE, Markland TE, Miller III TF. Ring-polymer molecular dynamics: Quantum effects in chemical dynamics from classical trajectories in a extended phase space. *Annu. Rev. Phys. Chem.* **64**, 387-413 (2013).
40. Craig IR, Manolopoulos DE. Chemical reaction rates from ring polymer molecular dynamics. *J. Chem. Phys.* **122**, 084106 (2005).
41. Craig IR, Manolopoulos DE. A refined ring polymer molecular dynamics theory of chemical reaction rates. *J. Chem. Phys.* **123**, 034102 (2005).
42. Collepardo-Guevara R, Craig IR, Manolopoulos DE. Proton transfer in a polar solvent from ring polymer reaction rate theory. *J. Chem. Phys.* **128**, 144502 (2008).
43. Collepardo-Guevara R, Suleimanov YV, Manolopoulos DE. Bimolecular reaction rates from ring polymer molecular dynamics. *J. Chem. Phys.* **130**, 174713 (2009).
44. Suleimanov YV, Allen JW, Green WH. RPMDrate: bimolecular chemical reaction rates from ring polymer molecular dynamics. *Comput. Phys. Comm.* **184**, 833-840 (2013).
45. Suleimanov YV, Aoiz FJ, Guo H. Chemical reaction rate coefficients from ring polymer molecular dynamics: Theory and practical applications. *J. Phys. Chem. A* **120**, 8488-8502 (2016).
46. Suleimanov YV. Surface diffusion of hydrogen on Ni(100) from ring polymer molecular dynamics. *J. Phys. Chem. C* **116**, 11141-11153 (2012).
47. Fang W, Richardson JO, Chen J, Li X-Z, Michaelides A. Simultaneous deep tunneling and classical hopping for hydrogen diffusion on metals. *Phys. Rev. Lett.* **119**, 126001 (2017).
48. Li C, Li Y, Jiang B. First-principles surface reaction rates by ring polymer molecular dynamics and neural network potential: role of anharmonicity and lattice motion. *Chem. Sci.* **14**, 5087-5098 (2023).
49. Zhang L, Zuo J, Suleimanov YV, Guo H. Ring polymer molecular dynamics approach to quantum dissociative chemisorption rates. *J. Phys. Chem. Lett.* **14**, 7118-7125 (2023).
50. Nitz F, *et al.* Thermal rates and high-temperature tunneling from surface reaction dynamics and first-principles. *J. Am. Chem. Soc.* **146**, 31538–31546 (2024).
51. Zhang L, Nitz F, Borodin D, Wodtke AM, Guo H. Ring-polymer molecular dynamics rates for hydrogen recombinative desorption on Pt(111). *Prec. Chem.* **in press**, (2025).
52. Torrie GM, Valleau JP. Non-physical sampling distributions in Monte Carlo free energy estimation: Umbrella sampling. *J. Comput. Phys.* **23**, 187-199 (1977).
53. Barducci A, Bussi G, Parrinello M. Well-tempered metadynamics: A smoothly converging and tunable free-energy method. *Phys. Rev. Lett.* **100**, 020603 (2008).
54. Santra AK, Goodman DW. Catalytic oxidation of CO by platinum group metals: from ultrahigh vacuum to elevated pressures. *Electrochim. Acta* **47**, 3595-3609 (2002).
55. Alavi A, Hu P, Deutsch T, Silvestrelli PL, Hutter J. CO oxidation on Pt(111): An ab initio density functional theory study. *Phys. Rev. Lett.* **80**, 3650-3653 (1998).
56. Zhou L, Kandratsenka A, Campbell CT, Wodtke AM, Guo H. Origin of thermal and hyperthermal $CO_2$ from CO oxidation on Pt surfaces: The role of post-transition-state dynamics, active sites, and chemisorbed $CO_2$. *Angew. Chem. Int. Ed.* **58**, 6916-6920 (2019).
57. Zhang Y, Hu C, Jiang B. Embedded atom neural network potentials: Efficient and accurate machine learning with a physically inspired representation. *J. Phys. Chem. Lett.* **10**, 4962-4967 (2019).

58. Zhang Y, Lin Q, Jiang B. Atomistic neural network representations for chemical dynamics simulations of molecular, condensed phase, and interfacial systems: Efficiency, representability, and generalization. *WIREs Comput. Mol. Sci.* **13**, e1645 (2023).
59. Luo S, Zhao Y, Truhlar DG. Improved CO adsorption energies, site preferences, and surface formation energies from a neta-generalized gradient approximation exchange–correlation functional, M06-L. *J. Phys. Chem. Lett.* **3**, 2975-2979 (2012).
60. Lininger CN*, et al.* Challenges for density functional theory: calculation of CO adsorption on electrocatalytically relevant metals. *Phys. Chem. Chem. Phys.* **23**, 9394-9406 (2021).
61. Li W-L*, et al.* Critical Role of Thermal Fluctuations for CO Binding on Electrocatalytic Metal Surfaces. *JACS Au* **1**, 1708-1718 (2021).
62. Ertl G, Neumann M, Streit KM. Chemisorption of CO on the Pt(111) surface. *Surf. Sci.* **64**, 393-410 (1977).
63. Eichler A. CO oxidation on transition metal surfaces: reaction rates from first principles. *Surf. Sci.* **498**, 314-320 (2002).
64. Verheij LK, Lux J, Anton AB, Poelsema B, Comsa G. A molecular beam study of the interaction of CO molecules with a Pt(111) surface using pulse shape analysis. *Surf. Sci.* **182**, 390-410 (1987).
65. Feibelman PJ*, et al.* The CO/Pt(111) puzzle. *J. Phys. Chem. B* **105**, 4018-4025 (2001).
66. Rohrdanz MA, Zheng W, Clementi C. Discovering mountain passes via torchlight: Methods for the definition of reaction coordinates and pathways in complex macromolecular reactions. *Annu. Rev. Phys. Chem.* **64**, 295-316 (2013).
67. Chen M. Collective variable-based enhanced sampling and machine learning. *Eur. Phys. J. B* **94**, 211 (2021).
68. Laio A, Parrinello M. Escaping free energy minima. *Proc. Natl. Acad. Sci. USA* **99**, 12562-12566 (2002).
69. Zhang Y, Hu C, Jiang B. Embedded atom neural network potentials: Efficient and accurate machine learning with a physically inspired representation. *J. Phys. Chem. Lett.* **10**, 4962-4967 (2019).
70. Hammer B, Hansen LB, Nørskov JK. Improved adsorption energetics within density functional theory using revised Perdew-Burke-Ernzerhof functionals. *Phys. Rev. B* **59**, 7413-7421 (1999).
71. Plimpton S. Fast parallel algorithms for short-range molecular dynamics. *J. Comput. Phys.* **117**, 1-19 (1995).
72. Tribello GA, Bonomi M, Branduardi D, Camilloni C, Bussi G. PLUMED 2: New feathers for an old bird. *Comput. Phys. Comm.* **185**, 604-613 (2014).

**Acknowledgements:**

This work was supported by Strategic Priority Research Program of the Chinese Academy of Sciences (XDB0450101 to B. J. and Z. L.), National Natural Science Foundation of China (22325304 and 22221003 to B. J., 21503130 and 11674212 to Y. L.). H.G. thanks the National Science Foundation (CHE-2306975 to H.G.) and the Alexander von Humboldt Foundation for support. Calculations were performed at the Supercomputing Center of USTC and Hefei Advanced Computing Center. We thank Dr. Dmitriy Borodin and Prof. Alec M. Wodtke for providing us the experimental data and thoughtful discussion.

**Author contributions:**

B. J. conceived the idea behind this work and led the supervision of this research. C. L. wrote the code and did all calculations with the initial help of X. Z.. C. L., H. G. and B. J. wrote the first draft of the manuscript and all authors edited its subsequent versions.

**Conflicts of interest:**

There are no conflicts to declare.

**Additional information:**

**Supplementary information**

The online version contains supplementary material.

TOC graphic:

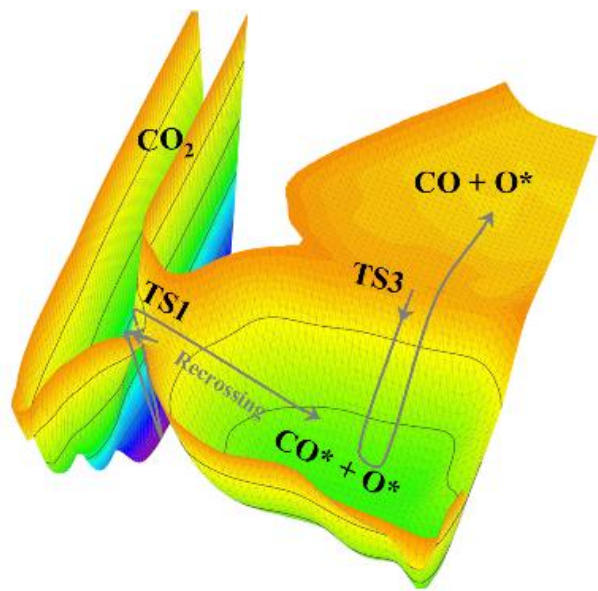

# Supplementary information for

# First Principles Reactive Flux Theory for Surface Reactions: Multiple Channels and Recrossing Dynamics

Chen Li[1], Xiongzhi Zeng[1], Yongle Li[2], Zhenyu Li[1], Hua Guo[3], and Bin Jiang[1,*]

[1]State Key Laboratory of Precision and Intelligent Chemistry, Department of Chemical Physics, University of Science and Technology of China, Hefei, Anhui 230026, China

[2]Department of Physics, International Center of Quantum and Molecular Structures and Shanghai Key Laboratory of High Temperature Superconductors, Shanghai University, Shanghai 200444, China.

[3]Department of Chemistry and Chemical Biology, Center for Computational Chemistry, University of New Mexico, Albuquerque, New Mexico 87131, USA

*: corresponding author: *bjiangch@ustc.edu.cn*

## S1. Computational Details

All density functional theory (DFT) calculations in this work were performed by the Vienna Ab initio Simulation Package (VASP)[1,2] with spin-polarization. The Pt(111) surface was modeled by a four-layer slab within a 3×3 surface unit cell, in which the top two layers are movable. The interslab distance is 20 Å. The truncated kinetic energy of 400 eV in the plane wave basis set and the 5×5×1 Monkhorst-Pack $k$-point mesh ensure the convergence of the properties of the stationary points. It is known that functionals with the generalized gradient approximation often predict incorrectly the favorable site of CO adsorption on the Pt(111) surface.[3] Fortunately, our focus here is on CO oxidation, with the description of CO adsorption of less concern. Specifically, the revised Perdew–Burke–Ernzerhof (RPBE) functional[4] was selected, which predicts the CO adsorption energy on Pt(111) reasonably well.[5-7]

The potential energy surface (PES) for CO oxidation on Pt(111) was constructed by the embedded atom neural network (EANN) method.[8-11] The EANN model decomposes the total energy into atomic energies that depend on the local chemical environment, expressed by the embedded atom densities (EADs). Each EAD descriptor is simply evaluated by the square of the linear combination of the contracted Gaussian-type orbitals (GTOs) located at neighbor atoms inside a cutoff sphere with a radius of 6 Å, serving as the input vector of each atomic NN to output the atomic energy. The GTO is expressed as,

$$\varphi\left(\mathbf{r}_{ij}\right) = x_{ij}^{l_x} y_{ij}^{l_y} z_{ij}^{l_z} \exp\left(-\alpha \left|r_{ij} - r_s\right|^2\right), \tag{S1}$$

where $\mathbf{r}_{ij}$ and $r_{ij}$ are the position vector of the central atom $i$ relative to the $j$th neighbor atom and its norm respectively, the hyperparameters $\alpha$ and $r_s$ determine the width and center of the Gaussian function, the total orbital angular momentum and its projection onto Cartesian axes ($L=l_x+l_y+l_z$) specify the spatial distribution of the GTO. Specifically, eleven radical functions and $L$ up to 2 were utilized to generate 33 EAD descriptors. Each atomic NN consisting of two hidden layers with 32 and 64 neurons. A total of 5035 DFT points with both energies and gradients were collected by an uncertainty driven active learning strategy[12, 13] to fit the EANN PES. These data points cover all relevant reaction pathways very well and include changes in the surface configuration. The root mean squared errors (RMSEs) with respect to the total energy and atomic gradients are 43.7 meV per cell (namely ~2.1 meV per mobile atom) and 95.3 meV/Å, respectively.

**S2. Tests on the Choices of Collective Variables**

Fig. S1 illustrates all possible dynamical processes that can be explored in the well-tempered metadynamics (WT-MetaD) simulations. These processes are primarily characterized by the two C-O distances ($r_1$ and $r_2$) and the height ($h$) of the carbon atom above the surface. To justify the choice of these coordinates as collective variables (CVs), we tested the influence of different CV choices on the calculated free-energy profile for CO oxidation. Initially, only the distance between the C atom and the O2 atom, $r_2$, was selected as the CV, as it is directly related to the CO oxidation. In this scenario, the PES along $r_2$ will be gradually flatten with the added Gaussian bias potentials during WT-MetaD simulations, by which the CO adsorbate can approach the O2 atom to form a $CO_2$ molecule. Since $CO_2$ desorption is almost barrierless on the PES (as shown in Fig. S1), the resulting $CO_2$ molecule tends to move up and down on the surface during the simulation, as depicted in Fig. S2a, rather than effectively sampling the RC for CO oxidation, $r_2$. To solve this issue, it is useful to include $h$ as the second CV. By adding Gaussians along $h$, the $CO_2$ moiety moves back to the surface and repeatedly accesses the $CO_2$ dissociation and CO+O recombination pathways. This facilitates more thorough sampling the CO oxidation channel ($r_2$), as shown in Fig. S2b. Interestingly, this treatment inevitably brings in the CO desorption channel. However, with this choice, the C-O1 bond ($r_1$) can be broken and the trajectory may become trapped in the equivalent (CO + O1) co-adsorption well (as depicted in the left side of Fig. S1). To address this problem, we choose to include all three CVs ($r_1$, $r_2$, and $h$), which allow the trajectory to traverse all relevant channels, sampling the phase space more adequately, as shown in Fig. S2c. We note that this is not the only possible choice, but it is intuitive to define CVs in Euclidean distances, which are required for using Eq. (7) in the main text.

To assess the influence of the selected CVs, we calculated the free-energy profile for CO oxidation when only one or two CVs are used, and compared them with the converged result obtained with the three CVs. Note that the free-energy profile is converged only when the CV(s) are sufficiently sampled. Fig. S2d shows that as the number of CVs increases, the free-energy profile of CO oxidation gradually converges to the result obtained using three CVs. This illustrates the necessity and accuracy of our approach, highlighting the importance of using an appropriate set of CVs to ensure proper sampling of the relevant phase space.

In addition, the CO oxidation free-energy profile was also calculated by umbrella sampling (US)[14] based on a single RC ($r_2$ here), which is a common choice in previous free-energy calculations.[15] In comparison, the US-based result deviates substantially from the converged free-energy profile obtained by WT-MetaD. Notably, the US method fails to capture the CO diffusion barrier between different surface sites. This discrepancy likely arises because the US algorithm is restricted to sampling only along the pre-defined RC, while the WT-MetaD algorithm does not have such a restriction and can explore diffusion paths in any direction.

**S3. Recrossing trajectories**

We note that the transmission coefficients are contributed by two different types of trajectories that experience recrossing in the reactive flux theory. For the CO oxidation, these two types of trajectories are analyzed in Fig. S3 showing their $h$ and $r_2$ distances, along with the total kinetic energy of the $CO_2$ moiety ($E_k$) and its component in the Z direction ($E_{kZ}$), as a function of time. Fig. S3a shows one of the dominant types of recrossing trajectories, accounting for about 90% of all recrossing trajectories observed. Clearly, after initial sampling at TS1, the trajectory moves to the product side to form the $CO_2$ molecule, in which the C*-O* bond is highly excited, evidenced by its large kinetic energy. However, the $CO_2$ translational energy along the Z-direction is so small that the nascent $CO_2$ molecule quickly recrosses the TS1 and returns to the reactant side (OC*+O*) instead of desorption.[16] This kind of recrossing trajectories start with positive velocities towards the product side ($CO_2$) and recross the TS1 to arrive the reactant side, so referred to as the route $R \to P \to R$. The remaining ~10% recrossing trajectories follow another route $P \to R \to P$, starting with negative velocities toward the reactant side and re-crossing TS1 to the product side, as shown in Fig. S3b.

For the CO desorption channel, we also find two types of recrossing behaviors. Different from the CO oxidation channel, however, now recrossing trajectories following the $\mathbf{P}\rightarrow\mathbf{R}\rightarrow\mathbf{P}$ route dominate, accounting for about 93% recrossing trajectories. One such exemplary trajectory is illustrated in Fig. S4a with the associated $h$ distance evolved as a function of time. In the process of the OC* species sliding from the TS3 to the surface, it is decelerated by the repulsive force from the surface, and perhaps more importantly from the adsorbed oxygen atom, then quickly bounces back to move away from the surface. Eventually, the CO recrosses the TS3 and returns to the vacuum. A minor fraction of recrossing trajectories follows the $\mathbf{R}\rightarrow\mathbf{P}\rightarrow\mathbf{R}$ route, which may initially travel a very short distance toward the vacuum, then recross the TS3 back and forth, and eventually move to the adsorption side, as illustrated in Fig S4b.

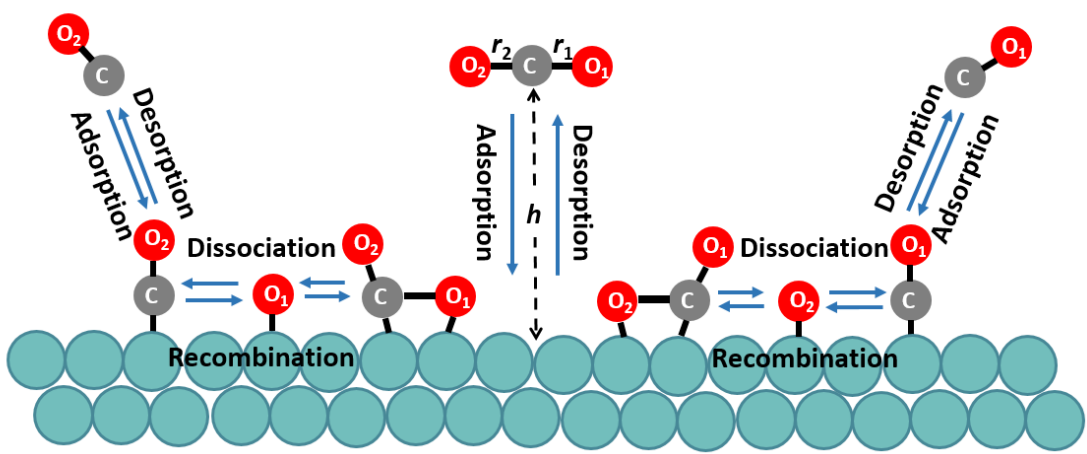

**Fig. S1.** Elementary steps in the CO oxidation process on the Pt(111) surface.

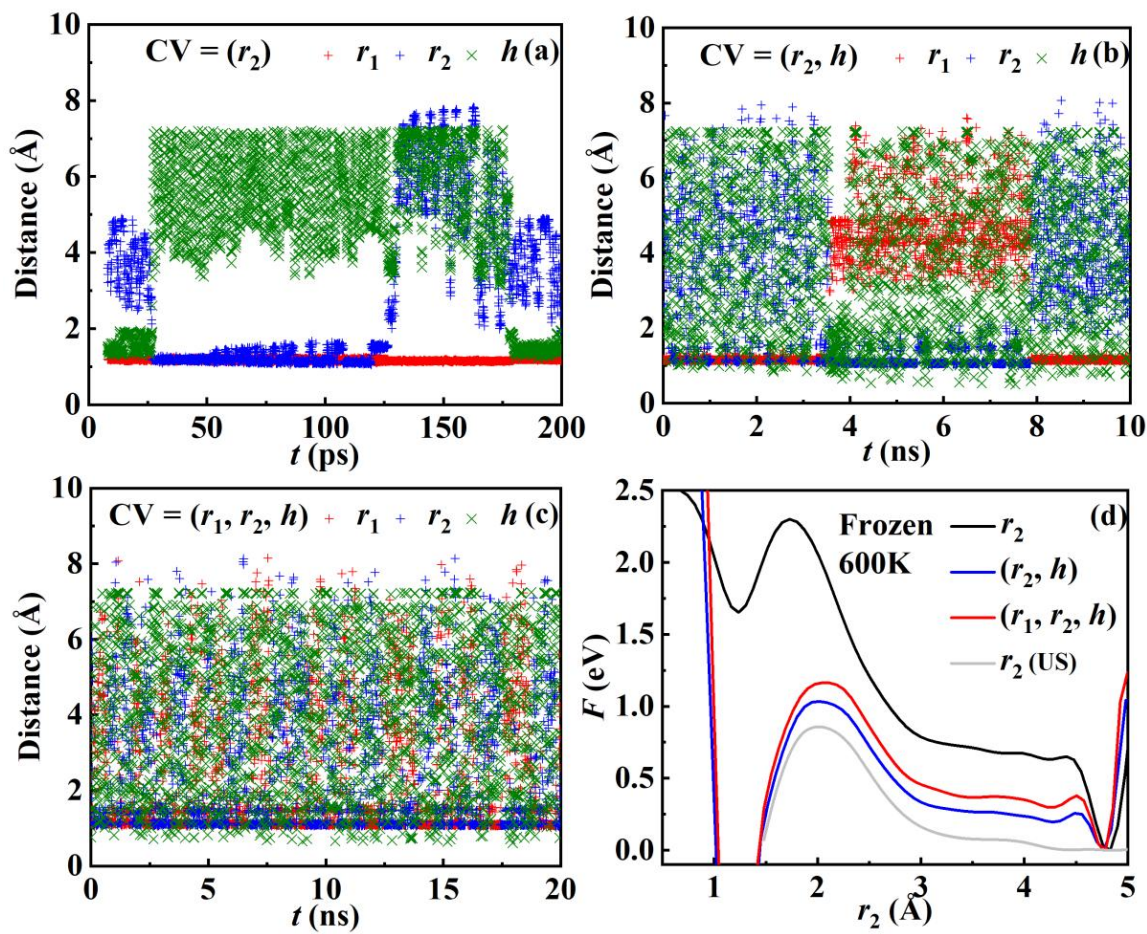

**Fig. S2. Exemplary trajectories in WT-MetaD simulations for different CV choices.** Evolution of $r_1$, $r_2$ and $h$ coordinates as a function of time for three exemplary trajectories with different choices of RCs, including **a** $r_1$, **b** $r_2$ and $h$, and **c** $r_1$, $r_2$ and $h$, respectively. Here WT-MetaD simulations are performed at 600 K and the Pt(111) surface is frozen. **d** Corresponding free-energy profiles of CO oxidation obtained by different choices of CVs and enhanced sampling schemes, see text.

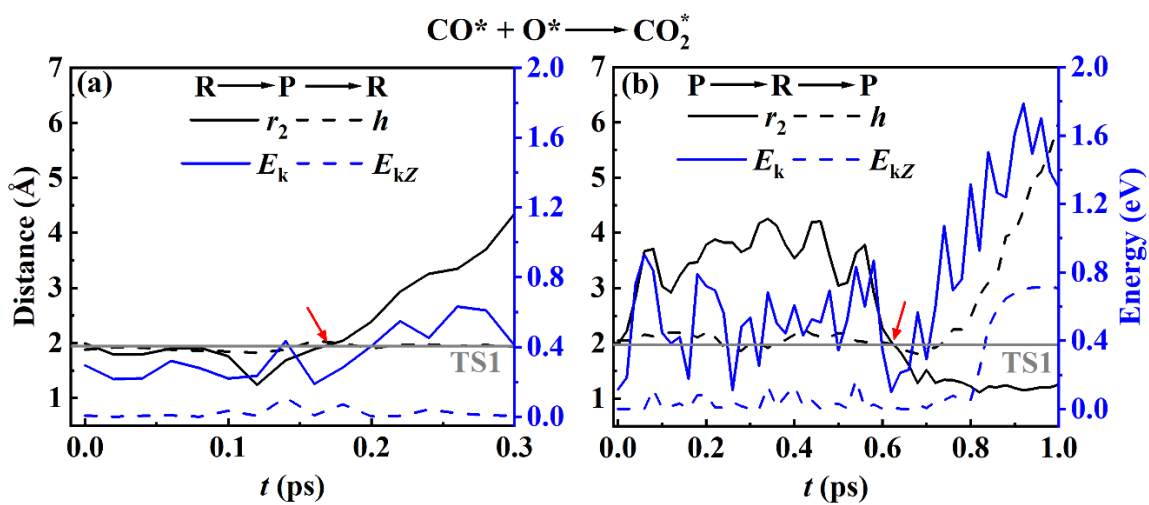

**Fig. S3.** Two types of recrossing trajectories for the CO oxidation at 600 K. **a** and **b** demonstrate $R \rightarrow P \rightarrow R$ and $P \rightarrow R \rightarrow P$ recrossing routes, respectively, where $r_2$ and $h$ distances, and total kinetic energy ($E_k$) and its component in $Z$ direction ($E_{kZ}$) are shown as a function of time. The $r_2$ value corresponding to the TS1 is represented by a gray line and the red arrows indicate where the recrossing occurs.

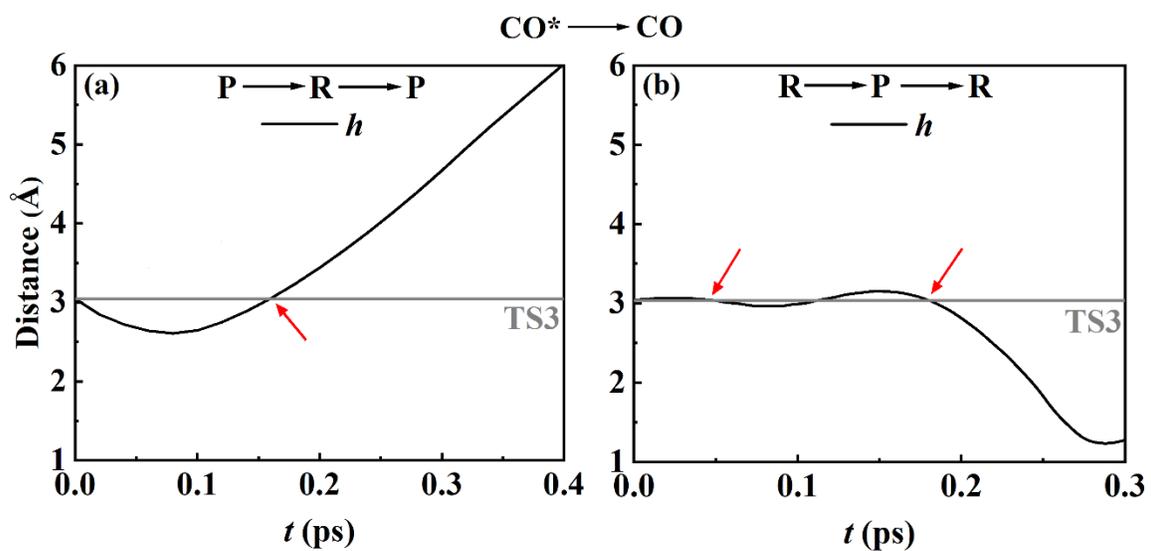

**Fig. S4.** Two types of recrossing trajectories for the CO desorption at 600 K. **a** and **b** demonstrate $P\rightarrow R\rightarrow P$ and $R\rightarrow P\rightarrow R$ recrossing routes, respectively, with their $h$ values varying as a function of time. The $h$ value corresponding to TS3 is represented by a gray line and the red arrows indicate where the recrossing occurs.

**Table S1: Comparison of geometric parameters of the stationary points along the reaction path for CO oxidation on Pt(111) predicted by DFT and the EANN PES (in brackets).**

| Species | $d_{C-Pt}$ (Å) | $r_1$ (Å) | $r_2$ (Å) | $\theta$ (deg) |
| --- | --- | --- | --- | --- |
| $CO_2$ | 7.00 (7.01) | 1.18 (1.18) | 1.18 (1.18) | 180.0 (180.0) |
| Linear $CO_2$ | 4.46 (4.78) | 1.18 (1.18) | 1.18 (1.18) | 180.0 (180.0) |
| TS2 | 2.34 (2.29) | 1.21 (1.21) | 1.24 (1.24) | 147.0 (145.3) |
| Bent $CO_2$ | 2.11 (2.12) | 1.22 (1.22) | 1.29 (1.29) | 133.3 (133.7) |
| TS1 | 1.94 (1.93) | 1.17 (1.17) | 1.96 (1.96) | 110.8 (110.7) |
| CO*+O* | 2.03 (2.03) | 1.19 (1.19) | 3.55 (3.55) | 97.5 (97.4) |
| CO+O* | 6.77 (6.78) | 1.15 (1.15) | 7.44 (7.43) | 139.2 (139.0) |

$d_{C-Pt}$, $r_1$ and $r_2$ are the distances between the C atom and the nearest Pt, O1 and O2 atoms, respectively. $\theta$ is the OCO bond angle.

**References:**

1. Kresse G, Furthmuller J. Efficient iterative schemes for ab initio total-energy calculations using plane wave basis set. *Phys. Rev. B* **54**, 11169-11186 (1996).
2. Kresse G, Furthmuller J. Efficiency of ab initio total energy calculations for metals and semiconductors using plane wave basis set. *Comp. Mater. Sci.* **6**, 15-50 (1996).
3. Feibelman PJ, *et al.* The CO/Pt(111) puzzle. *J. Phys. Chem. B* **105**, 4018-4025 (2001).
4. Hammer B, Hansen LB, Nørskov JK. Improved adsorption energetics within density functional theory using revised Perdew-Burke-Ernzerhof functionals. *Phys. Rev. B* **59**, 7413-7421 (1999).
5. Luo S, Zhao Y, Truhlar DG. Improved CO adsorption energies, site preferences, and surface formation energies from a neta-generalized gradient approximation exchange–correlation functional, M06-L. *J. Phys. Chem. Lett.* **3**, 2975-2979 (2012).
6. Lininger CN, *et al.* Challenges for density functional theory: calculation of CO adsorption on electrocatalytically relevant metals. *Phys. Chem. Chem. Phys.* **23**, 9394-9406 (2021).
7. Li W-L, *et al.* Critical Role of Thermal Fluctuations for CO Binding on Electrocatalytic Metal Surfaces. *JACS Au* **1**, 1708-1718 (2021).
8. Zhang Y, Hu C, Jiang B. Embedded atom neural network potentials: Efficient and accurate machine learning with a physically inspired representation. *J. Phys. Chem. Lett.* **10**, 4962-4967 (2019).
9. Zhang Y, Hu C, Jiang B. Accelerating atomistic simulations with piecewise machine-learned ab Initio potentials at a classical force field-like cost. *Phys. Chem. Chem. Phys.* **23**, 1815-1821 (2021).
10. Zhang Y, Xia J, Jiang B. Physically motivated recursively embedded atom neural networks: Incorporating local completeness and nonlocality. *Phys. Rev. Lett.* **127**, 156002 (2021).
11. Zhang Y, Xia J, Jiang B. REANN: A PyTorch-based end-to-end multi-functional deep neural network package for molecular, reactive, and periodic systems. *J. Chem. Phys.* **156**, 114801 (2022).
12. Lin Q, Zhang Y, Zhao B, Jiang B. Automatically growing global reactive neural network potential energy surfaces: A trajectory-free active learning strategy. *J. Chem. Phys.* **152**, 154104 (2020).
13. Lin Q, Zhang L, Zhang Y, Jiang B. Searching Configurations in Uncertainty Space: Active Learning of High-Dimensional Neural Network Reactive Potentials. *J. Chem. Theory Comput* **17**, 2691-2701 (2021).
14. Torrie GM, Valleau JP. Non-physical sampling distributions in Monte Carlo free energy estimation: Umbrella sampling. *J. Comput. Phys.* **23**, 187-199 (1977).
15. Xu J, Huang H, Hu P. An approach to calculate the free energy changes of surface reactions using free energy decomposition on ab initio brute-force molecular dynamics trajectories. *Phys. Chem. Chem. Phys.* **22**, 21340-21349 (2020).
16. Zhou L, Kandratsenka A, Campbell CT, Wodtke AM, Guo H. Origin of thermal and hyperthermal $CO_2$ from CO oxidation on Pt surfaces: The role of post-transition-state dynamics, active sites, and chemisorbed $CO_2$. *Angew. Chem. Int. Ed.* **58**, 6916-6920 (2019).